\documentclass[aps,twocolumn,prb,superscriptaddress]{revtex4}

\usepackage{epsfig}
\usepackage{graphicx}
\usepackage{amssymb,amsmath}
\usepackage{dcolumn}
\usepackage{bm}
\usepackage{psfrag}
\usepackage{yfonts}
\usepackage{version}
\usepackage{natbib}
\usepackage{longtable}

\newcommand{\cref}[1]{chapter~\ref{#1}}

\newcommand{\Cref}[1]{Chapter~\ref{#1}}

\newcommand{\bra}[1]{\langle\,{#1}\, |}
\newcommand{\ket}[1]{|\,{#1}\,\rangle}

\newcommand{\V}{V}
\newcommand{\ElTransE}{\epsilon}

\newcommand{\Kern}{Q}

\newcommand{\om}{\omega}

\usepackage{color}
\usepackage{ulem}  
\begin{document}

\title{Photonics meets excitonics: natural and artificial molecular aggregates}

\author{Semion K. Saikin}
\email{saykin@fas.harvard.edu}
\affiliation{Department of Chemistry and Chemical Biology, Harvard University, 12 Oxford Street, Cambridge, MA 02138, USA}
\affiliation{Department of Physics, Kazan Federal University,18 Kremlyovskaya Street, Kazan 420008, Russian Federation}

\author{Alexander Eisfeld}
\email{eisfeld@mpipks-dresden.mpg.de}
\affiliation{Max-Planck-Institut f\"ur Physik komplexer Systeme, N\"othnitzer Str.~38, D-01187 Dresden, Germany}

\author{St\'ephanie Valleau}
\affiliation{Department of Chemistry and Chemical Biology, Harvard University, 12 Oxford Street, Cambridge, MA 02138, USA}

\author{Al\'an Aspuru-Guzik}
\email{aspuru@chemistry.harvard.edu}
\affiliation{Department of Chemistry and Chemical Biology, Harvard University, 12 Oxford Street, Cambridge, MA 02138, USA}


\begin{abstract}
Organic molecules store the energy of absorbed light in the form of charge-neutral molecular excitations -- Frenkel excitons. Usually, in amorphous organic materials, excitons are viewed as quasiparticles, localized on single molecules, which diffuse randomly through the structure. However, the picture of incoherent hopping is not applicable to some classes of molecular aggregates -- assemblies of molecules that have strong near field interaction between electronic excitations in the individual subunits. Molecular aggregates can be found in nature, in photosynthetic complexes of plants and bacteria, and they can also be produced artificially in various forms including quasi-one dimensional chains, two-dimensional films, tubes, etc. In these structures light is absorbed collectively by many molecules and the following dynamics of molecular excitation possesses coherent properties. This energy transfer mechanism, mediated by the coherent exciton dynamics, resembles the propagation of electromagnetic waves through a structured medium on the nanometer scale. The absorbed energy can be transferred resonantly over distances of hundreds of nanometers before exciton relaxation occurs. Furthermore, the spatial and energetic landscape of molecular aggregates can enable the funneling of the exciton energy to a small number of molecules either within or outside the aggregate. In this review we establish a bridge between the fields of photonics and excitonics by describing the present understanding of exciton dynamics in molecular aggregates.
\end{abstract}

\maketitle

\section{Introduction}

Advances in nanotechnology supported by our understanding of material properties on the microscopic level persistently drive the field of photonics to the nanometer scale. With the development of photonic crystals \cite{Yablonovitch_RPL87, Photonic_crystals_Nat97} and semiconductor optical cavities \cite{Vahala_Nat03, Khitrova_NatPhys06} the size of optical devices, usually limited by light diffraction in free space, can been scaled down to hundreds of nanometers just by exploiting the material's dielectric constant. Advances in the design of plasmonic metamaterials \cite{QuLeKr98_1331_,Kr03_210_,Ebbesen_Nat03} permitted this limit to be pushed even further down to the order of tens of nanometers. The natural question which arises is:  what will be the next limit and how can it be approached?

\begin{figure}[b]
  \begin{center}
   \includegraphics[width=0.95\linewidth]{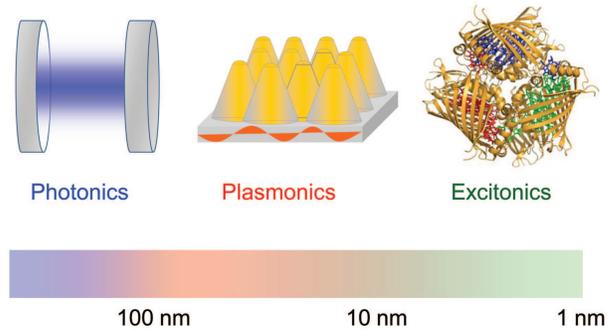}
  \caption{Schematic illustration of the length scales characterizing the different physical phenomena present the different fields discussed in this review.}
    \label{fig:scales}
  \end{center}
\end{figure}

In this review we consider one of possible pathways towards reaching the goal of optical devices on the nanometer scale: using molecular aggregates as photon processing elements on the true nanometer scale, see Fig.~\ref{fig:scales}. Molecular aggregates are assemblies of molecules held in place by non-covalent interactions. These single molecules mostly keep their nuclear and electronic structure. Examples are molecular crystals as well as nanoscale self-assembled structures, molecular films and light harvesting systems in photosynthesis. If the lowest allowed electronic transitions in the composing molecules are within the visible part of the optical spectrum and if the molecules have large absorption and fluorescence cross-sections, the interaction between molecular electronic transitions is strong enough to transfer the light absorbed from one molecule to the others, via a resonant de-excitation/excitation process.

This mechanism is closely related to the near-field energy transfer between classical dipoles (antennas) \cite{Ho25_722_,Fr30_198_,De64_393_,Ph73_227_,ZiSi10_144107_,BrEi11_051911_}. In this sense, the exciton dynamics in molecular aggregates
resembles the propagation of light in a metamaterial where dye molecules play the role of functional elements. In contrast to excitons in inorganic semiconductors, where free charge carrier mobility determines the exciton transport, in aggregates, electrons remain localized on molecules while the excitation is transferred.

 A canonical example of molecular aggregates that possess coherent exciton properties are J-aggregates. These are aggregates of fluorescent molecules, discovered about 80 years ago independently  by Scheibe \cite{Scheibe1936,Sc37_212_} and Jelley \cite{Jelley1936}. They observed that the formation of aggregates is accompanied by drastic changes in the optical properties. The broad absorption line corresponding to the molecular excitation is shifted to the red side of the spectrum and becomes much narrower.
 Later, these types of molecular aggregates were named J-aggregates after Jelley. J-aggregates represent only one class of molecular aggregates. For instance, there exist aggregates where the absorption line is blue-shifted, so called H-aggregates. Some aggregates exhibit several J-bands \cite{KiDa06_20363_} and some have both J-and H-bands \cite{No77_151_}. In aggregates that consist of non-equivalent monomers even more complicated absorption structures can be found \cite{AmVaGr00__}.

Molecular aggregates also appear as functional units in nature. For instance, they form the absorbing and energy transferring parts of light-harvesting complexes in plants and some types of bacteria and algae \cite{AmVaGr00__}. In these complexes the exciton absorbed by the antenna aggregate has to be funnelled to the reaction center  -- the part where the exciton energy is used to create free charges to be employed in a chemical reaction. The efficiency and robustness of light absorption and exciton transfer in light harvesting complexes may be crucial for the survival of photosynthetic organisms under evolutionary pressure.

Shortly after the discovery of the self-assembled organic dye aggregates the close connection of these artificial structures to the natural light harvesting systems was recognized \cite{Sc37_795_,FrTe38_861_} and the exciton model of Frenkel \cite{Fr30_198_,Fr31_17_} has been used to explain the observed changes in optical properties and the transfer of energy along the aggregate \cite{FrTe38_861_}.

 In Frenkel's exciton theory, which is based on the classical resonance interaction theory of Holtsmark \cite{Ho25_722_}, the electronic excitation in the aggregate is not confined to a single monomer, but it is coherently delocalized over many monomers in the form of ``excitation waves''. From superpositions of these excitation waves ``excitation packets'' can be formed, which describe the coherent motion of (localized) excitations \cite{Fr31_17_}. Already in these early works it was established that coupling to internal and external vibrational modes, imperfections of the aggregates and disorder induced by the environment strongly influences the ``coherence size'' of the exciton waves and modifies the optical and transport properties.

When the interaction with the environment (internal vibrations, solvent, etc.) is much stronger than the resonant excitation transfer interaction, the excitation becomes more or less localized on one monomer and the transfer is no longer described by coherent exciton motion but it becomes an incoherent hopping process. F\"orster derived an elegant formula for the rate constants for transport of excitation from one monomer to the other \cite{Foe48_55_,Foe65_93_}. This rate is proportional to the overlap of the donor emission spectrum and the absorption spectrum of the acceptor molecule and depends on the inverse sixth power of the distance between donor and acceptor. Typically, in molecular aggregates, one is in a regime in-between these two extreme cases, i.e.\ the transport is neither fully coherent nor incoherent. This complicates the theoretical modeling and the interpretation of experiments.

The remarkable optical and transport properties of molecular aggregates have led to a variety of applications. Right from the beginning, molecular aggregates were employed as wavelength selective sensitizers in photography \cite{Ta96__,TaSuKa06_16169_}.
Recent applications include the measurement of membrane potentials \cite{ReSmCh91_4480_} or the  design of colorants \cite{HoRi01_4331_}.
Some molecular aggregates form self-assembled supramolecular flexible fluorescent
fibers \cite{HiKeVa96_4049_} which may have applications in thin-film optical and optoelectronic devices, for example by employing optical bistability of J-aggregates \cite{MaGlFe00_1170_}.
Molecular aggregates can be also utilized for sensing applications. For instance, it has been demonstrated that a large absorption cross section combined with fast exciton diffusion may be used to enhance fluorescence from a small number of dye molecules adsorbed  or embedded in an aggregate film \cite{Kuhn_JAP88, Akselrod_ASCNano12}. Due to fast exciton diffusion within the aggregate the excitons explore the aggregate. Once they find a molecule with an electronic transition close to the J-band the exciton can be transferred inelastically to the adsorbed molecule. Moreover, dye aggregates might play an important role in the development of efficient, low-cost  artificial light harvesting units (like organic solar cells)\cite{Dae02_81_}.

This review gives a brief, yet by no means complete, overview of our experimental and theoretical understanding of excitons in molecular aggregates. Most of the aspects in this work are covered at the advanced introductory level, and for a deeper understanding we refer to more detailed studies such as \cite{Wurthner2011,Da62__,AmVaGr00__,Ko96__,ScRu06_683_,Loch_Rev2011}. The review is structured as follows:
In Sec.~\ref{sec:Basic_properties}, we describe basic excitonic properties of molecular aggregates starting with single molecules and molecular dimers and then introducing several examples of artificial and natural molecular aggregates. In Sec.~\ref{sec:Exp_charcterization}, we overview the main experimental techniques that allow for probing the structural properties of molecular aggregates and excitation dynamics. Section \ref{sec:Models} introduces the main theoretical approaches utilized for modeling of excitons in aggregates. Section \ref{sec:Hybrid_structures} shows how molecular aggregates can be combined with photonic and plasmonic structures. Finally, we conclude the review in Section \ref{sec:Conclusions}.

\section{Basic properties of molecular aggregates}
\label{sec:Basic_properties}

\subsection{Properties of monomers}
\begin{figure}
  \begin{center}
   \includegraphics[width=0.7\linewidth, clip=true]{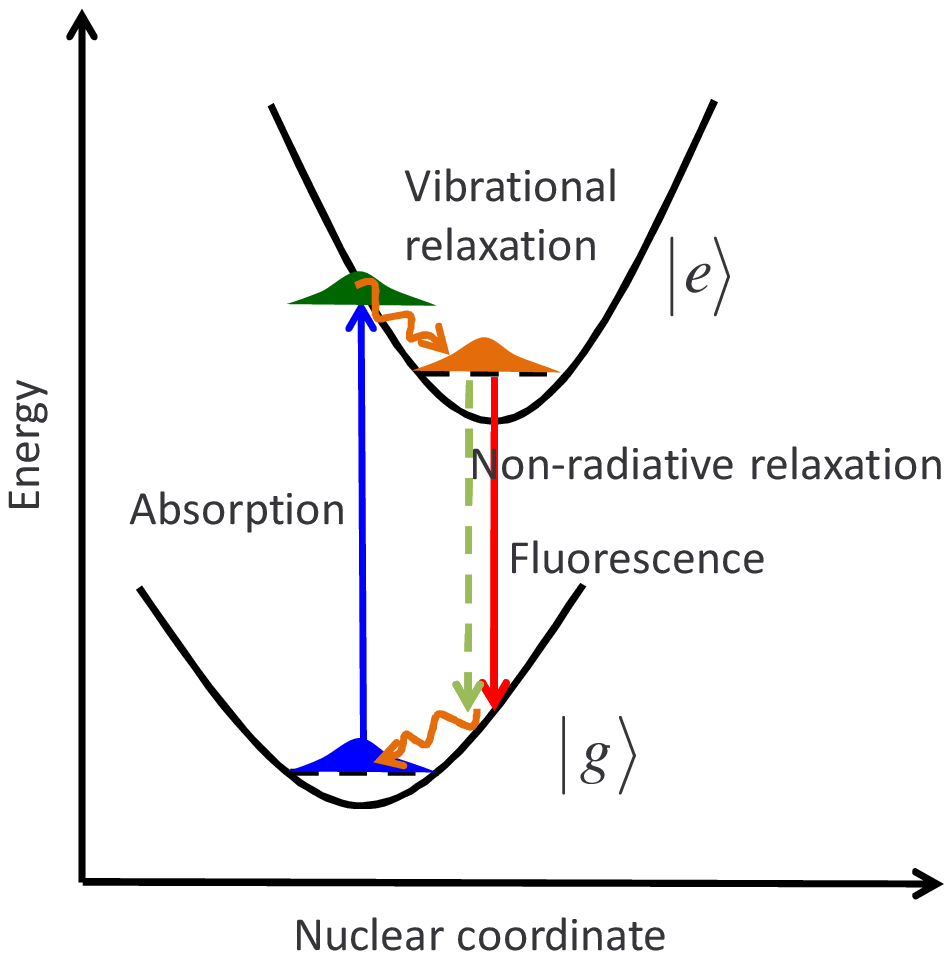}
  \caption{Excitation/de-excitation processes in a single molecule illustrated on a two-level molecular energy diagram. The parabolic surfaces correspond to the ground state $|g\rangle$ and the first excited $|e\rangle$ electronic states of the molecule. The horizontal axis indicates the displacements of nuclei from their equilibrium positions. The molecule, initially in the ground electronic and vibrational states, is excited by absorbing light -- blue arrow. During the absorption process the positions of the nuclei are not changed. The light absorption also induces molecular vibrations. Due to the interaction with the environment the molecular vibrations are equilibrated and the molecule relaxes to the bottom of the excited electronic state -- orange wavy arrow. Finally, the molecule relaxes down to the ground electronic state by emitting a photon -- red arrow (fluorescence), or without photon emission -- green dashed arrow (non-radiative process). In the ground state the induced vibrations are also equilibrated due to interaction with the environment.}
    \label{fig:molec_spectrum}
  \end{center}
\end{figure}

Molecules in the aggregates largely retain their electronic and nuclear structure. Thus, it is natural to start our discussion with individual molecules interacting with light. The interaction of a single molecule with light is schematically illustrated in Fig.~\ref{fig:molec_spectrum}. If the molecule has an optically allowed electronic transitions within the photon spectrum it can be excited by absorbing light at the corresponding frequency.  The transition time between the ground and excited state of the molecule is of the order of tens of attoseconds -- the time that a single photon interacts with a molecule. During this short time interval the positions of the nuclei in the molecule are not changed. Initially the molecule is in its ground state geometry, i.e.\ in a minimum of the electronic ground state potential $|g\rangle$ indicated by the blue wave-packet in Fig~\ref{fig:molec_spectrum}. The equilibrium positions of atoms for the excited states, usually, are different from those of the ground state. Therefore, after the transition to the state $|e\rangle$ the molecule is in a transient non-equilibrium state where molecular vibrational modes are also excited (green wave-packet in Fig~\ref{fig:molec_spectrum}). Then, due to the interaction of the molecule with its environment the molecule relaxes towards the energetic minimum of the excited state, i.e.\ the excited vibrational modes are relaxed (wavy arrow and orange wavepacket in Fig~\ref{fig:molec_spectrum}). This relaxation of the molecular geometry can also be accompanied by a rearrangement of the environmental molecules to a configuration with lower total energy. At ambient conditions this process occurs over a timescale of hundreds of femtoseconds to several picoseconds. Finally, on timescales of tens of picoseconds to nanoseconds the molecule relaxes to its electronic ground state either emitting a photon (fluorescence) or transferring energy to other degrees of freedom, for instance vibrations (non-radiative relaxation \cite{MeOs95__}).
Usually, the absorption and emission spectra exhibit a progression of peaks stemming from the coupling to vibrational modes with high energy ($\sim$150 meV). These peaks are broadened by the same order of magnitude, due to coupling to a multitude of low energy modes of the molecule and the surroundings. Emission takes place from the thermally relaxed excited state, which is typically located in the low energy wing of the absorption spectrum. The relaxation energy is related to the energy difference between the maxima of the absorption and emission spectra, the so-called Stokes shift.

The absorption efficiency of a particular electronic transition can often be characterized by the corresponding transition dipole - a matrix element of a dipole operator between the ground and the excited molecular states
\begin{equation}
 \vec{d} = \langle e|q \cdot \vec{r}|g\rangle.
 \label{transition dipole}
\end{equation}
Here and in the following we use an arrow symbol to indicate three-dimensional vectors in real space.
 The dipole operator is characterized by its charge $q$ and the position operator $\vec{r}$.
The transition dipole is not a measurable value and is defined up to a complex phase factor (if magnetic interactions can be neglected, which is often the case, then the wavefunction can be chosen real and $\vec{d}$ becomes a real vector). The absorption and emission strengths of the respective transition are proportional to $|\vec{d}|^2$. In the absorption spectra the transition frequency is determined by the energy difference $\varepsilon$ between the ground and excited state.

\begin{figure}
  \begin{center}
   \includegraphics[width=0.8\linewidth, clip=true]{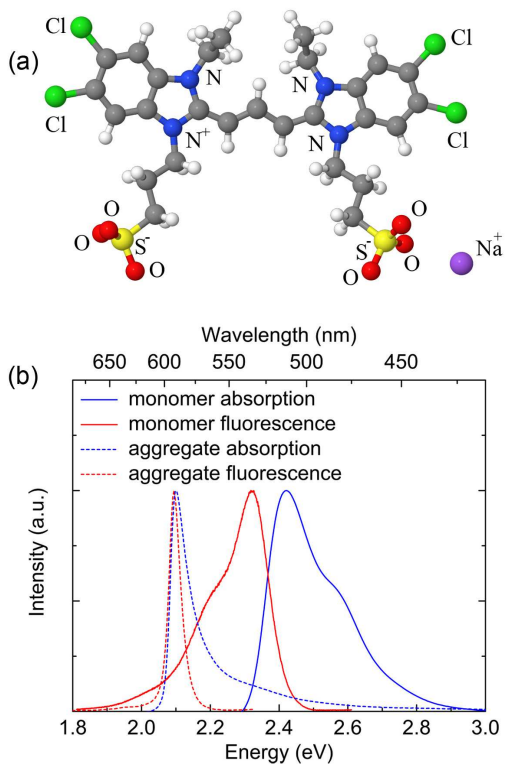}
  \caption{TDBC  fluorescent dye that forms 1D and 2D molecular aggregates. (a) Molecular structure, grey and white spheres represent carbons and hydrogens respectively. (b) Normalized absorption and fluorescence spectra of TDBC molecules taken in a solution 
  (solid lines) and in a 2D aggregated form (dashed lines).}
    \label{fig:TDBC_monomer}
  \end{center}
\end{figure}

\begin{figure}
  \begin{center}
   \includegraphics[width=0.8\linewidth, clip=true]{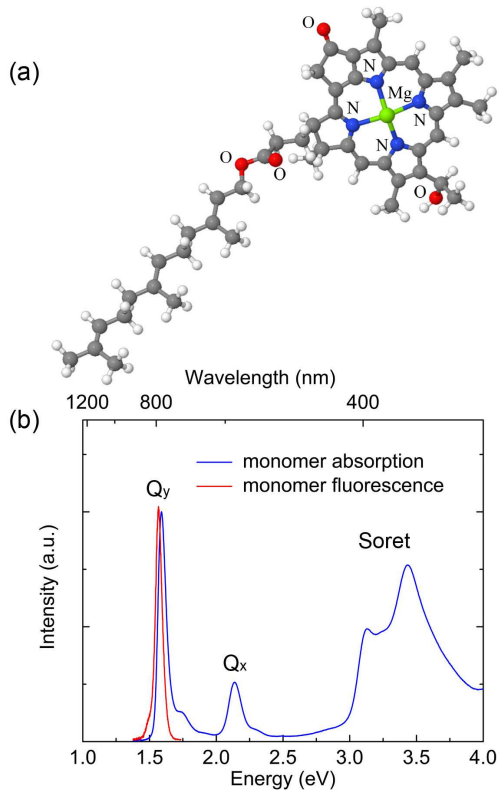}
  \caption{(a) Bacteriochlorophyll (Bchl) molecule - light absorbing pigment that is a basic structural element of light-harvesting complexes in phototrophic bacteria. Its structure is similar to chlorophyll - the photosynthetic pigment in plants.; (b) Absorption and fluorescent spectra of Bchl molecules. }
    \label{fig:Bchl_monomer}
  \end{center}
\end{figure}

Structures and optical spectra of representative molecules are shown in Figs.~\ref{fig:TDBC_monomer} and \ref{fig:Bchl_monomer}.
In Fig.~\ref{fig:TDBC_monomer} we show a TDBC molecule, which can be used as an illustrative example for a broader class of cyanine dyes. Beside the monomeric absorption and  fluorescence spectra (in solution), spectra of aggregates formed on a glass substrate are also shown. The lowest electronic transition in TDBC is characterized by a large transition dipole of the order of $|\vec{d}|\approx 10$~Debye (Debye units are commonly used for molecular dipoles, $1$~Debye~$\approx 0.208\,$e$\cdot$\AA\, where $e$ denotes the electron charge), which is aligned with the backbone of the molecule. The higher excited states in TDBC are well separated from the lowest one, which indicates that, for low intensity optical absorption and exciton dynamics it is sufficient to take only two electronic states into account.
The absorption spectrum of the molecule in a solution shows a broad line in the green part of the spectrum with a partially resolved vibronic structure.  Note also that the emission from a monomer is roughly a mirror-image of the absorption,  where the maximum is shifted to lower energies. This is the Stokes shift, which is caused by the relaxation/reorganisation after an electronic transition. The mirror image of absorption and fluorescence indicates that the vibrational frequencies  and their coupling to electronic transitions are very similar in the electronic ground and excited state. Upon aggregation the absorption and fluorescence lines become much narrower with no structure and are shifted to the orange-red color.
The Stokes shift in the aggregated form is also much smaller, which indicates a reduced coupling to the environment. Aggregates composed of TDBC have, for example, been produced in solution \cite{KiDa06_20363_} and on surfaces \cite{Bradley2005}. By changing the side groups different geometrical arrangements can be achieved \cite{KiDa06_20363_}.

Another representative example of molecules forming aggregates is the bacteriochlorophyll (Bchl)-- a pigment molecule, which is a functional element of photosynthetic systems in phototrophic bacteria, see Fig.~\ref{fig:Bchl_monomer}(a). The structure of Bchl is similar to that of chlorophyll - the photosynthetic pigment in plants. Both are derivatives of porphyrin, compelexed with Mg$^{2+}$.  The lowest electronic transition, which is involved in the energy transfer, is the so called $Q_y$-band. The corresponding transition dipole lies in the plane of the porphyrin ring and is about $|\vec{d}| \approx 5$~Debye. The transition dipole corresponding to the next electronic state is denoted as $Q_x$.  It is perpendicular to the $Q_y$ dipole, and the associated transition strength is much smaller.  The Soret band which lies in the ultraviolet has stronger absorption but is usually not involved in the energy transfer. As in the case of TDBC the emission occurs from the lowest excited state and the emission spectrum is roughly a mirror image of the absorption band corresponding to this state.

\subsection{The aggregate}
Organic dye molecules can self-aggregate in different types of structures. Sometimes the  aggregation is driven by electrostatic forces pushing molecules to adsorb on a surface; sometimes hydrophobic parts of the molecules repel the water and tend to collect together such as in synthetic tube aggregates \cite{KiDa06_20363_}.

\begin{figure}[h]
  \begin{center}
   \includegraphics[width=0.7\linewidth, clip=true]{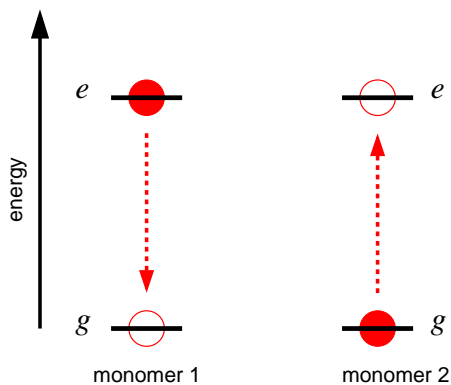}
  \caption{Illustration of the excitation transfer due to resonant near-field interaction between the state $|e\rangle|g\rangle$  and $|g\rangle|e\rangle$.}
    \label{fig:resonant_transfer}
  \end{center}
\end{figure}

A specific  property of molecular aggregates with strong
inter-monomer couplings is that the absorption and fluorescent
spectra substantially differ from  the spectra of the molecules
which form the aggregates while the electronic structure of the
molecules is not modified. The intermolecular distance in an
aggregate is large enough that the electron tunneling between
different molecules can be neglected. Therefore, only the Coulomb
interaction between the electronic transitions in monomers is
responsible for the spectral modifications and the intermolecular
energy transfer. This interaction is similar to the non-radiative
near-field coupling between plasmonic structures, which is
mediated by virtual photons.  It turns out that the major
contribution of the Coulomb interaction in the spectra of
aggregates is from the excitation transfer via resonant
transitions. We will refer to this main interaction term between
the monomers as F\"orster coupling. The F\"orster coupling, to a
good approximation, can be described by restricting to dipole
transitions only (see below Eq.~\ref{eq:dip-dip-term}). The
interaction is visualized in Fig.~\ref{fig:resonant_transfer} for
a situation in which initially monomer 1 is electronically excited
and monomer 2 is in the electronic ground state. Then, in a
resonant process molecule 1 is deexcited and simultaneously the
second monomer is excited. While this time-ordered discrete
picture is easy for visualisation in reality the exciton energy is
transferred coherently.
For large distances between the monomers the interaction between monomers 1 and 2 can be approximated by the dipole-dipole term
\begin{equation}
\label{eq:dip-dip-term}
V_{12}= \frac{1}{R_{12}^3}
\Big(\vec{d}_1\cdot\vec{d}_2- 3 (\vec{d}_1 \cdot \hat{R}_{12})(\vec{d}_2 \cdot \hat{R}_{12})
\Big),
\end{equation}
where ${R}_{12}$ denotes the distance between the monomers, $\hat{R}_{12}$ is the corresponding direction vector, and the vectors $\vec{d}_n$  are the transition dipoles introduced in Eq.~(\ref{transition dipole}).

If the intermolecular distance is comparable with the size of the molecules, then the interaction between the two monomers can no longer be described  adequately using the point dipole-dipole interaction of Eq.~(\ref{eq:dip-dip-term}). Often, it is then sufficient to replace the point-dipoles in Eq.~(\ref{eq:dip-dip-term}) by extended dipoles \cite{Czikklely1970}. For even higher accuracy more elaborate schemes have been developed (see e.g.~\cite{KrScFl98_5378_}).

\subsubsection{The dimer}
\begin{figure}
  \begin{center}
   \includegraphics[width=0.9\linewidth, clip=true]{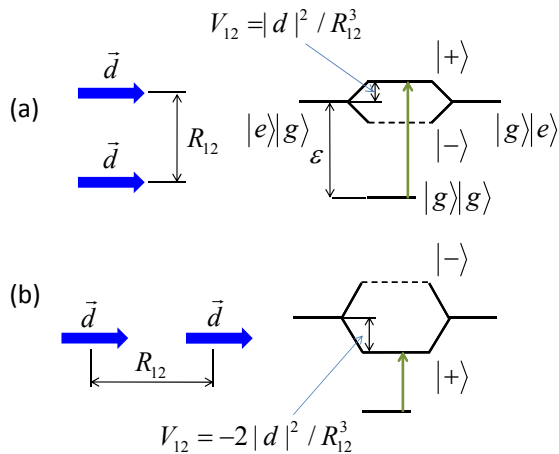}
  \caption{Interaction between electronic transitions dipoles (blue arrows) of two molecules. The directions of the arrows reflect the relative phase of the transition dipoles. For the considered molecular alignments only one of the optical transitions is allowed for the dimer (green arrows). (a) The optically allowed transition of the dimer is shifted to the blue part of the spectrum, which is similar to H-aggregation; (b) the optically allowed transition is shifted to the red, which correspond to J-aggregation.}
    \label{fig:dimer}
  \end{center}
\end{figure}

Before discussing general molecular aggregates, let us describe some basic results for the case of two coupled identical monomers, ignoring nuclear degrees of freedom for the moment. The eigenstates of the uncoupled monomers can be taken as a basis. These states are  $|g\rangle|g\rangle$,  $|e\rangle|g\rangle$, $|g\rangle|e\rangle$,  $|e\rangle|e\rangle$.
While the states $|g\rangle|g\rangle$ and   $|e\rangle|e\rangle$ are still suitable to describe the ground state and the doubly excited state of the dimer, respectively, the two degenerate single exciton states $|e\rangle|g\rangle$ and  $|g\rangle|e\rangle$ are no longer eigenstates of the coupled system, because of the resonant transfer interaction.
The corresponding Hamiltonian is written as
$H=\varepsilon_1|1\rangle\langle 1|+ \varepsilon_2|2\rangle \langle 2| + V_{12}(|1\rangle \langle 2|+ |2\rangle\langle 1|)$ (we choose the energy of the monomer ground state as zero), where we have introduced the shorthand notation
$|1\rangle\equiv|e\rangle|g\rangle$ and $|2\rangle\equiv|g\rangle|e\rangle$
or equivalently in a matrix representation
\begin{equation}
H=\left(
\begin{array}{cc}
\varepsilon_1 & V_{12}\\
V_{12} & \varepsilon_2
\end{array}
\right).
\end{equation}
For identical energies of excited states $\varepsilon_1=\varepsilon_2\equiv \varepsilon$ the eigenstates of the dimer are superpositions $|\pm\rangle = \frac{1}{\sqrt{2}}\big(|e\rangle|g\rangle \pm |e\rangle|g\rangle \big)$ where the electronic excitation is coherently delocalized over both monomers. The corresponding eigenenergies are $\ElTransE_{\pm}=\varepsilon \pm V_{12}$. Note that the magnitude and sign of the interaction depends sensitively on the distance and orientation of the two monomers. Fig.~\ref{fig:dimer} illustrates how the relative orientations of molecular transition dipoles change the frequency of electronic excitations in a dimer.  If the transition frequencies of the involved monomers are equal and the transition dipoles are parallel, only the transition between the ground and the $|+\rangle$ state is optically allowed. It is detuned by the energy $V_{12}$ from that of the non-interacting monomers.
From Eq.~(\ref{eq:dip-dip-term}) one can see that for case (a), where the transition dipoles are orthogonal to the distance vector between the monomers one has positive $V_{12}\sim |d|^2/R_{12}^3$. For the case (b) the interaction is negative with $V_{12}\sim -2 |d|^2/R_{12}^3$. Therefore, the alinement of the molecules shown in Fig.~\ref{fig:dimer}~(a) and (b) will be seen as a blue and red shift of the absorption line, respectively. As it has been noted before, the states with only one excited monomer such as $|g\rangle|e\rangle$ and $|e\rangle|g\rangle$ are not eigenstates of the system anymore. Thus, if one of these state is populated initially, the energy will oscillate between them back and forth coherently.

Let us briefly mention that the dimer often appears as a first step of aggregation \cite{KoHaKa81_498_}  and thus has been investigated intensively (see e.g.~\cite{FuGo64_2280_,KReMa96_99_,SeLoWue07_6214_,Ei07_321_,WoMo09_15747_,GuZuCh09_154302_}) . This interest in the dimer is also due to the fact that many of the more realistic models for aggregates where the environment and the internal vibrations are included can only be solved efficiently for the dimer.

\subsubsection{Arbitrary aggregates}
The theoretical description of the dimer can easily be extended to arbitrarily arranged monomers.
As in the case of the dimer the total ground state is taken as a product of states where all monomers are in the ground state $\ket{g_{\rm el}^{\rm agg}}=\ket{g}_1\cdots  \ket{g}_N$.
We are interested in the properties of aggregates in the linear absorption regime. Therefore, it is sufficient to take into account only the states with at most one electronic excitation on the aggregate,
such as
\begin{equation}
  \ket{n}\equiv \ket{g}_1\cdots\ket{e}_n\cdots \ket{g}_N,
  \label{eq:pi_n}
\end{equation}
in which monomer $n$ is electronically excited and all other monomers are in their electronic ground state.
In this {\it one-exciton manifold} approximation the Hamiltonian can be written as
\begin{equation}
\label{eq:H_elec}
H^e=\sum_{n=1}^N\varepsilon_n \ket{n}\bra{n}+\sum_{nm}V_{nm}\ket{n}\bra{m}
\end{equation}
where $V_{nm}$ is the Coulomb dipole-dipole interaction, see Eq.~\ref{eq:dip-dip-term}, which transfers excitation from monomer $n$ to $m$.

\begin{figure}
  \begin{center}
   \includegraphics[width=0.7\linewidth, clip=true]{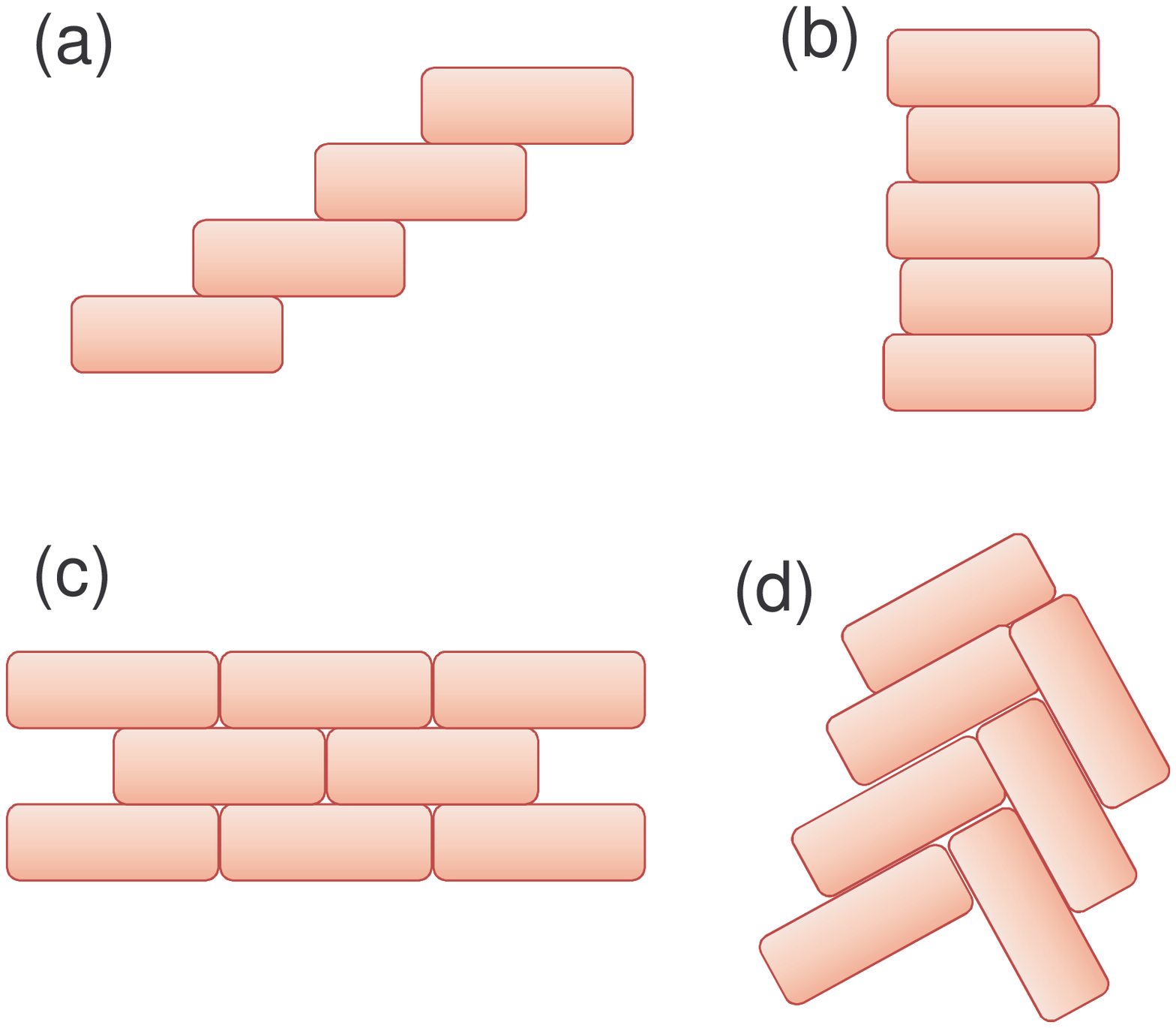}
  \caption{Illustrations of 1D and 2D planar molecular aggregates. Each block corresponds to a monomer forming the aggregate. (a) - staircase and (b) - ladder models for 1D packing; (c) - brickstone  and (d) - herringbone 2D packing models.}
    \label{fig:chain_and_film}
 \end{center}
\end{figure}

Examples of commonly discussed 1D and 2D planar structures are shown in Fig.~\ref{fig:chain_and_film}. While 1D models are well studied theoretically (see e.g.~Refs.~\cite{MoFiKi57_723_,Ho59_343_,Ho66_208_,Bi70_4987_,Kn84_73_,ScFi84_269_,Sp91_3424_,Kn93_8466_,HoScFr00_73_}) and are frequently used to characterize excitons in self-assembled molecular aggregates, it is not clear whether ideal 1D molecular chains are formed in experiments \cite{Wurthner2011}. Usually, they are subsystems of higher dimensional structures such as films or crystals.

 The 1D model of molecular aggregates is convenient because many results can be obtained analytically \cite{MaKue00__,AmVaGr00__}.
For a perfect very long chain with $N$ molecules  the eigenstates are well described by ``exciton waves''
\begin{equation}
\label{eigVal}
\ket{\phi_{j}}= \frac{1}{\sqrt{N}}\sum_{n=1}^N e^{i \frac{2 \pi}{N} j n} \ket{n}
\end{equation}
with the corresponding eigenenergies
\begin{equation}
\label{eigE}
E_{j}=\ElTransE+ 2V \cos \!\Big(\frac{2 \pi j }{N}\Big).
\end{equation}
In the last equation, for the sake of simplicity, we have considered the interactions between nearest neighbors only, which are denoted by $V$.
A discussion of finite one-dimensional aggregates can be found e.g.\ in Refs.~\cite{MaKue00__,AmVaGr00__}.

For the perfect linear chain with parallel monomers, the largest transition dipole correspond to the excitation of the state with a minimal number of nodes. In the case of J-aggregation, Fig.~\ref{fig:chain_and_film}(a), this state is at the bottom of the exciton spectrum shifted by about several hundreds of meV from the monomer transition while for H-aggregates, Fig.~\ref{fig:chain_and_film}(b), this shift is to the blue part of the spectrum. In reality, to compare exciton transitions in aggregates with excitations in single molecules one also has to include Van der Waals interaction with off-resonant excitations \cite{AmVaGr00__,Ei07_321_} and the vibronic structure of molecular excitation \cite{AmVaGr00__,Ei07_321_,EiBr06_376_}. However, the simplified model presented above qualitatively describes the exciton states in aggregates.

2D molecular aggregates can be formed, for instance, as Langmuir-Blodgett films \cite{PaGaHi08_5946_}, deposited molecule-by-molecule on a surface \cite{PrNiDi05_165207_,Bradley2005}, or by spin coating \cite{MiOnMi93_577_}. The structure of the aggregate is determined by the molecular properties such as their shape, charge, etc., and the assembly method. For instance, for cyanine dyes \cite{MiBeBe00_1973_} (TDBC is one example of the class of molecules) some commonly assumed aggregation structures are brickstone, Fig.~\ref{fig:chain_and_film}(c), and herringbone, Fig.~\ref{fig:chain_and_film}(d).

3D molecular aggregates with translational symmetry are usually known as molecular crystals. Frequently, in these structures is possible to identify lower dimensional subsystems with a preferred interaction between molecular transition dipoles \cite{ScWo06__}. These directions would determine anisotropy of exciton transport. While on average the exciton transport in a crystal can be incoherent one may expect specific directions with large coherent exciton delocalization.

The above mentioned structures are only very simple examples of molecular aggregates. For instance, when driven by hydrophobic and hydrophilic forces molecules can aggregate in two-layer tubes of about 10 nanometers in diameter \cite{KiDa06_20363_,Dorthe_2012} or one can use templates like polypeptides or DNA to induce helical structures \cite{StBl61_1411_,SeCoKu99_2987_}.

\begin{figure}
  \begin{center}
   \includegraphics[width=0.8\linewidth, clip=true]{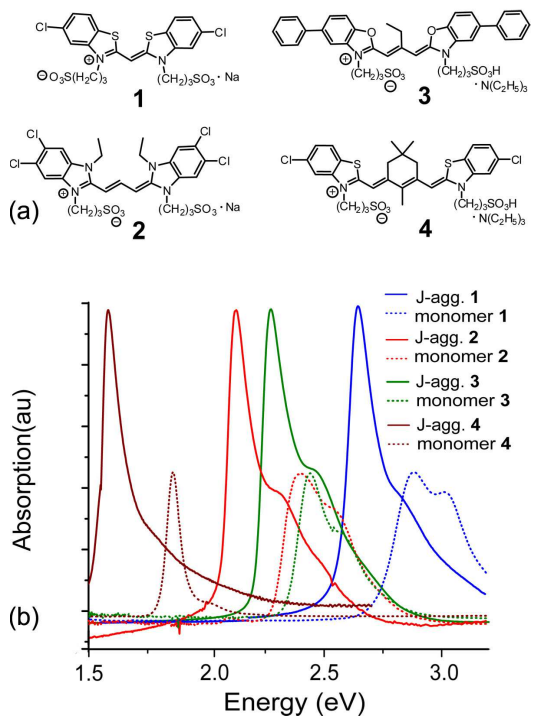}
  \caption{Absorption spectra of four different  cyanine dyes in monomeric and aggregated forms illustrating the variability in the aggregate optical properties. Adapted with permission from Ref.~\cite{Walker_NanoLett}. Copyright 2011 American Chemical Society. }
    \label{fig:dyes}
  \end{center}
\end{figure}

Because of the huge variety of organic dyes that can aggregate it is possible to create narrow J-band absorption  within an arbitrary part of visible and near IR spectrum, see Fig~\ref{fig:dyes}.

\subsection{Natural aggregates}
\begin{figure}
  \begin{center}
   \includegraphics[width=0.7\linewidth, clip=true]{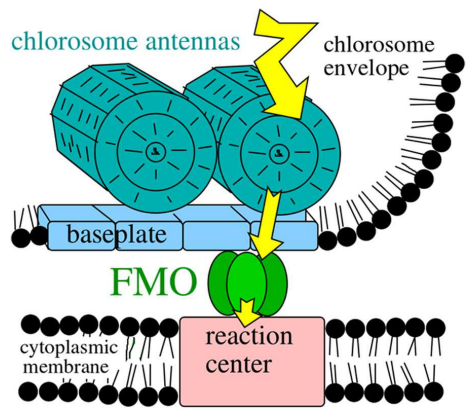}
  \caption{Schematic structure of light-harvesting complex in green sulfur bacteria. The solar light is absorbed by chlorosome antenna aggregates of Bchl molecules. Then, the energy in the form of excitons is transferred through the baseplate and Fenna-Matthews-Olson (FMO) protein complexes to the reaction center, where the charge separation occurs.}
    \label{fig:GSB}
  \end{center}
\end{figure}

In nature, photosynthetic complexes of plants, phototrophic bacteria and algae are composed of molecular aggregates that efficiently transfer excitons. As an example, in Fig.~\ref{fig:GSB} we sketch the energy transfer along the light harvesting system of  green sulfur bacteria (GSB).  These GSB can be found e.g.\ in deep sea. By developing a sophisticated light-harvesting antennae structure these organisms adapted to survive at very low light intensities \cite{Overmann1992, Beatty2005}. Energy absorption and transfer in GSB goes through a network of Bchl molecules, Fig.~\ref{fig:Bchl_monomer}, aggregated in several types of functional structures -- chlorosome antenna, baseplate, Fenna-Matthews-Olson protein complex (FMO), and the reaction center.

\begin{figure}
  \begin{center}
   \includegraphics[width=0.85\linewidth, clip=true]{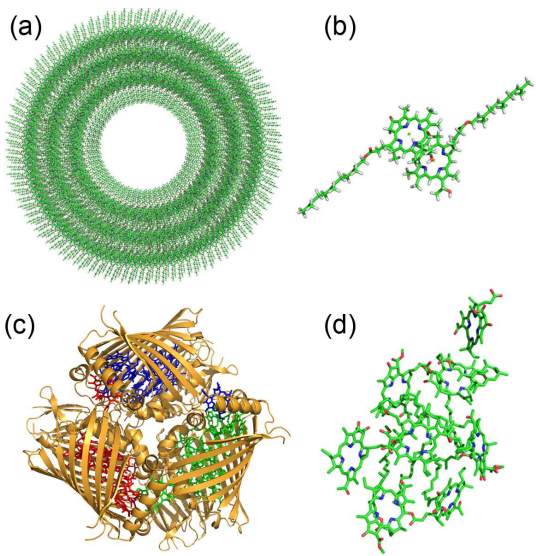}
  \caption{Examples of molecular aggregates in the light-harvesting complex of green sulfur bacteria. (a) Chlorosome antenna complex -- a cylindrical or ellipsoidal aggregate of Bchl molecules, several concentric aggregates are enclosed; (b) A Bchl dimer -- a unit block of the antenna complex \cite{Ganapathy2009}; (c) Fenna-Matthews-Olson (FMO) protein complex -- an excitonic ``wire''; (d) organization of Bchls in a monomer unit of FMO complex \cite{Tronrud09} }
    \label{fig:GSB_aggregates}
  \end{center}
\end{figure}

The photons are absorbed by the chlorosome -- an organelle which is bound to the bacterial membrane and has a size of several hundreds of nanometers. The chlorosome contains a disordered array of cylindrical or ellipsoidal molecular aggregates -- antennas. While different types of geometrical and structural disorders complicate the molecular structure characterization of natural aggregates, recent NMR analysis of mutant chlorosome antennas \cite{Ganapathy2009} suggested that Bchl molecules in them are arranged in an array of concentric helical structures illustrated in Fig.~\ref{fig:GSB_aggregates}(a). The F\"orster coupling between nearest-neighbor Bchl molecules in the chlorosome antenna complexes is of the order of 100~meV and the proposed molecular arrangement of the mutant should result in the formation of a J-band. Both experimental \cite{Dostal_2012}  and theoretical studies \cite{Taka2012}, show that the exciton spreads over a single antennae on the timescale of hundreds of femtoseconds.

Another functional element of GSB light-harvesting structure, is the Fenna-Matthews-Olson (FMO) complex, a protein which is depicted in Fig.~\ref{fig:GSB_aggregates}(c). FMO plays the role of a molecular wire transferring energy from the chlorosome to the reaction center. It is a trimer containing 8 Bchl molecules in each subunit. Unlike in the chlorosome antenna, Bchl molecules in FMO are held together by a protein cage, and also the F\"orster coupling between the molecules is several times weaker. However, from the multiple experimental and theoretical studies it is suggested that the exciton states in FMO are partly delocalized \cite{MiBrGr10_257_,ReMa98_4381_,AdRe06_2778_}.

%
\section{Experimental characterization of excitonic systems}
\label{sec:Exp_charcterization}

Much effort is devoted to  obtain the microscopic structure of the various aggregates (which in turn are needed for theoretical models), including the arrangement of the monomers, their spacial and energetic disorder, etc. Beside these conformational aspects there is also large interest in the dynamic properties of excitons such as exciton relaxation and dephasing rates, diffusion coefficients, etc.

X-ray crystallography can be used to obtain lattice properties of molecular aggregates, if the latter either possess an intrinsically long-range order (like molecular crystals or two-dimensional monolayers on substrate \cite{MIkSo11_1090_})  or can be crystallized. For instance, crystallization of photosynthetic light-harvesting complexes \cite{Olson_2004} had a large impact on the understanding of natural photosynthesis \cite{AmVaGr00__}. However, most  molecular aggregates cannot be crystallized. In recent years Cryo-Transmission electron microscopy has provided valuable information on the geometrical structure of many aggregates \cite{BeBoeOu00_5255_,VlAuSt09_2273_,Oostergetel_2007, Dorthe_2012}.

The classical way to obtain information on the geometry of the aggregate is by optical spectroscopy (in particular absorption, linear dichroism and circular dichroism) combined with theoretical modeling. For instance, the positions, intensities, polarizations and shapes of the aggregate absorption bands provide information about the strength of intermolecular coupling as well as relative orientations of the monomers \cite{AmVaGr00__,Ei07_321_,VlAuSt09_2273_}. Circular dichroism spectra can be used to analyze chirality of the structures, for instance, in studies of J-aggregate helices and tubes \cite{No77_151_,SpDaeOu00_8664_,DiKlKn02_11474_,DiPuHa04_14976_,EiKnBr07_104904_}.
Similarly, the emission spectra can be used to obtain information on the aggregate structure. These optical experiments are often performed under varying environmental conditions like solvent, temperature, concentration of monomers, alignment of the aggregates or pressure \cite{ReWi97_7977_,LiCh96_5359_,KaLeFr08_112004_}.
 More detailed information including separation of homogeneous and inhomogeneous line widths and femtosecond exciton dynamics can be obtained from non-linear 2D spectra \cite{Stiopkin06,Dostal_2012}.
Beside optical spectroscopy of electronic excitations a multitude of various experimental techniques is used to characterize aggregates, e.g.\ electroabsorption \cite{LuKr03_155_}, Fourier transform infrared (FTIR) spectroscopy  \cite{IlBa00_9331_}, nuclear magnetic resonance NMR \cite{RoStMu01_1587_} and Raman spectroscopy \cite{LeWueKe04_10284_}.

 With optical spectroscopy one probes the structure of energetic levels in the aggregate and phase relations between corresponding electronic transitions, from which one can infer the excitation dynamics in the aggregate. This might be a viable way for small systems, where the exciton is delocalized along the aggregate. However, one would like to follow the dynamics of the exciton also in real space, that means to measure the time-dependent probability to find excitation on a certain monomer.
Here in particular, diffusion constants are of interest, which characterize the spreading of the exciton after the initial coherences have died out.
Exciton diffusion coefficients cannot be measured directly using existing experimental techniques. The indirect methods include quenching of photoluminescence \cite{Lyons2005,Lunt2009}, photocurrent response \cite{Ghosh1978}, transient grating \cite{Salcedo1978} and exciton-exciton annihilation \cite{MoHaBr00_8847_,Shaw2010a,Akselrod2010,Lochbrunner2011}. Among the state-of-the-art techniques one could mention the recently developed coherent nanoscopy \cite{Aeschlimann2011}, which would allow for spatial resolution comparable with the exciton diffusion length.

\section{Models of exciton dynamics}
\label{sec:Models}

While the simple electronic model from Section~\ref{sec:Basic_properties} (see in particular  Eq.~(\ref{eq:H_elec})) already allows us to understand basic properties of molecular aggregates (for instance, the position of the J-band) it is not sufficient to describe important features such as the narrowing of the J-band (in particular in contrast to the broad H-band with vibronic structure). Also, many transport properties, e.g. dependence on temperature \cite{Scheblykin_chapter}, cannot be explained.
To this end it is necessary to include vibrational modes and the influence of the environment into the description. This can be done in various ways and there exists a multitude of theoretical models in the literature where each is best suited for a particular question and/or situation. For example, the F\"orster rate theory \cite{Foe65_93_} works well when the timescale of the transfer is very long compared to decoherence and vibrational relaxation of the excitation. In this Section we review the most common models.

\subsection{Static disorder}
Since usually the experiments are performed on ensembles of aggregates where each aggregate experiences a slightly different environment, it is common to average over different ``configurations'' of the aggregate. Most often the influence of the environment is treated simply as inducing random shifts of the transition frequencies of the monomers. For early works see e.g.~\cite{ScTo82_1528_,Kn84_73_,FiKnWi91_7880_}.
Often, it is assumed that each monomer sees ``its own'' local environment and the shifts induced by this local environment are uncorrelated from the shifts of the neighboring monomers and disorder of the coupling is ignored. This assumption could be called ``uncorrelated diagonal disorder''. Sometimes also the positions or couplings of the monomers are treated as random variables (off-diagonal disorder)\cite{FiKnWi91_7880_,MaDo99_255_}.
For each disorder realisation the exciton states are typically no longer delocalized over the full aggregate as in a perfect chain. This localisation becomes stronger for larger values of disorder \cite{Kn84_73_,Ma91_873_,FiKnWi91_7880_}. In the literature, the influence of various forms of correlations has been discussed, see e.g.~\cite{Kn84_73_,DoMa99_61_,WaEiBr08_044505_}. To account for temperature a variant of the open system model described below is adapted \cite{HeMaKn05_177402_,BeMaKn03_217401_,VlMaKn07_154719_}, where the environment of the disordered chain can lead to scattering and relaxation between the localized exciton states.
Such a model has been successfully used to describe optical and transfer properties of some aggregates \cite{ToMi96_2963_,GaPiSc00_3918_,BeMaKn03_217401_}. In particular at low temperatures the increase of mobility with increasing temperature could be explained.

In the above formulation the influence of internal vibrations of the monomers is  neglected. This seems to be a good approximation for the J-band where one can show that the electronic excitation is to a large extend decoupled from vibrational modes \cite{BrHe71_865_,ScFi84_269_,WaEiBr08_044505_}.

\subsection{Dynamic disorder, Haken-Strobel-Reineker model}

Fluctuations of the monomer transition frequencies or the intermolecular couplings that are fast compared to the exciton transfer time-scale are usually referred to as dynamic disorder.
Dynamical fluctuations play a dual role in exciton transport \cite{Wolynes1987, Wolynes1990, Rebentrost2009ENAQT}. In structures with large static disorder excitons are localized. Dynamic fluctuations in this case remove localization. However, if the fluctuations are strong this results in the dynamical localization of the exciton.
A seminal model to treat these fast fluctuations is due to Haken, Strobl and Reineker \cite{Haken1972,Haken1973,ReKr82__}, where the dynamical fluctuations are described by real stochastic processes, $\varepsilon_n(t)$ and $V_{nm}(t)$ and modeled using the density matrix \cite{Haken1972} or stochastic Schr\"odinger \cite{Stephanie_JCP12} equations. In the original work the stochastic processes have been chosen as white noise, i.e.\ delta-correlated in time. This model has been shown to capture coherent and incoherent excitation transfer on the same footing. Moreover, one should notice that the equation describing Haken-Strobel-Reineker (HRS) model for 1D systems is equivalent to telegrapher's equation, which further reflects the similarity between the exciton propagation in molecular aggregates and a wave propagation in a medium.

 The  Haken-Strobl-Reineker model, with several extensions, has been utilized to study exciton dynamics in natural and artificial molecular aggregates \cite{MaPo77_1424_,ReKr82__,PlHu08_113019_,Rebentrost2009ENAQT,EiBr12_046118_,Stephanie_JCP12}. The drawback of this model is that the exciton population does not relax to thermal equilibrium. There have been many investigations and extensions which have included colored noise and tried to cure the thermalization issue, e.g.~\cite{Ca85_101_,RaKnKe79_197_}. The original model also does not include the effect of strongly coupled high frequency modes of the monomers which is essential to describe the vibronic structure present in many H-aggregates. The recent development of so called quantum state diffusion methods \cite{DiSt97_569_,RoEiWo09_058301_} can be considered as an extension of the HSR model which cures these deficiencies at the cost of colored complex noise.

\begin{figure*}
  \begin{center}
   \includegraphics[width=0.95\linewidth, clip=true]{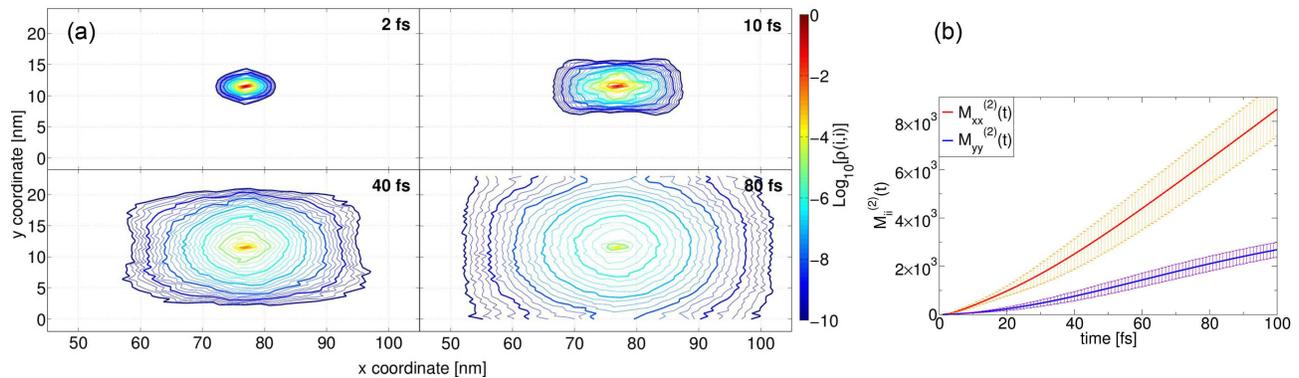}
  \caption{(a) Time dynamics of exciton wave function on a 2D lattice of TDBC molecules subject to static disorder (linewidth $\sigma=70\,\mathrm{meV}$) and dynamic fluctuations (linewidth $\Gamma = 30\,\mathrm{meV}$) of molecular electronic transitions. At time zero the exciton is localized in the center of the lattice. The contour plots show the population of lattice sites at four different times. (b)Second moments of the wave function corresponding to the exciton dynamics shown in (a). Two different transport regimes can be observed: the ballistic or a wave-like propagation on timescales of $20-30$~fs ($M^{(2)} \sim t^2$) and the diffusive motion ($M^{(2)} \sim t$) at longer times. Reprinted with permission from Ref.~\cite{Stephanie_JCP12}. Copyright 2012, American Institute of Physics.}
    \label{fig:HRS}
  \end{center}
\end{figure*}

Fig.~\ref{fig:HRS}(a) illustrates how the HRS model can be applied to exciton propagation on a brickstone lattice of TDBC dyes at ambient conditions \cite{Stephanie_JCP12}. The second moment of the exciton wavefunction, Fig.~\ref{fig:HRS}(b) shows that the initial propagation on the time scale of $20-30$~fs is ballistic or  wave-like, which is associated with the quadratic dependence of the second moment on time. For the longer timescale the time dependence is linear, which characterizes a diffusive motion.

\subsection{The vibronic model}

Let us first consider a single monomer, where only the two lowest electronic states are taken into account, see Fig.~\ref{fig:molec_spectrum}.
In the Born-Oppenheimer approximation \cite{Azumi_1977} we write the nuclear Hamiltonian of the electronic ground and excited states as
\begin{equation}
\label{HamMonGround}
H^{g/e}_n=\frac{1}{2} \sum_j (P_{nj}^2+ U_n^{g/e}(Q_{nj}))
\end{equation}
where $Q_{nj}$ are the (mass-scaled) nuclear coordinates of
monomer $n$ and  $P_{nj}$ are the corresponding momenta.
$U^{g/e}(Q_{nj})$ is the Born-Oppenheimer potential in the
electronic ground/excited state (see Fig.~\ref{fig:molec_spectrum}).

Then, for a molecular aggregate  the Hamiltonian in the electronic ground state is given by
$
H^g=\left(\sum_{n=1}^N H_n^g \right)\ket{g_{\rm el}}\bra{g_{\rm el}}.
$
We denote by $V_{nm}$ the transition dipole-dipole interaction between monomer $n$ and $m$, which, for simplicity, is taken to be independent of nuclear coordinates.  The Hamiltonian in the one-exciton manifold, Eq.~(\ref{eq:pi_n}), is then given by
\begin{equation}
\begin{split}
\label{HamTotExc}
H^e=\sum_{n=1}^N W_n\ket{n}\bra{n}
&+ \sum_{n,m=1}^N\V_{nm}\ket{n}\bra{m}.
\end{split}
\end{equation}
with the ``collective'' Born-Oppenheimer surfaces $W_n=H_n^e  +\sum_{m\ne n}^N H_m^g $.
Note that the structure of Eq.~(\ref{HamTotExc}) is very similar to the purely electronic Hamiltonian (\ref{eq:H_elec}). The only difference is that now the energies $\epsilon_n$ are replaced by operators for the nuclear motion. Upon transfer of excitation the nuclear wavepacket, according to Fig.~\ref{fig:molec_spectrum}, is no longer in its equilibrium position. Thus the excitation transfer is linked to excitation of nuclear motion. This vibrational model is sometimes combined with the static disorder approach \cite{Sp05_235208_,WaEiBr08_044505_,DiSpBo12_69_}.
The vibrational model offers a clear idea regarding why the J-band does not exhibit vibrational structure and the H-band does.

The coupling to vibrations usually slows down the propagation of the exciton, however, it can help to overcome energetic barriers caused by differences in the electronic transition energies. In this way efficient directed transport along a biased chain can be achieved.

Often it is justified to approximate the Born-Oppenheimer potentials, Eq.~(\ref{HamMonGround}), as harmonic potentials where the surface of the electronically excited state is just shifted relative to that of the ground electronic state, i.e.\ we have
$
H^g_n=\frac{1}{2}\sum_{j=1}^M(P_{nj}^2+\om_{nj}^2 Q_{nj}^2)
$
and for the excited electronic state
$
H_n^e=\ElTransE_{n}+\frac{1}{2}\sum_{j=1}^M\big(P_{nj}^2+\om_{nj}^2
(\Kern_{nj}- \Delta Q_{nj})^2\big)
$
where $\ElTransE_{n}$ denotes the energy difference between the
minima of the upper and lower potential energy surface and the
vibrational frequency of mode $j$ is denoted by $\omega_{nj}$ and
the shift between the minima of the excited and ground state
harmonic potential is denoted by
 $\Delta Q_{nj}$. The potential surfaces are sketched in Fig.~\ref{fig:molec_spectrum} for the case of one mode.
This harmonic approximation is widely employed in the literature, e.g.~\cite{Me64_445_,PeGo67_1019_,ScFi84_269_,EiBrSt05_134103_,SeLoWue07_6214_,SeWiRe09_13475_,DiSpBo12_69_}.

\subsection{Open system approaches}
Open system approaches are closely related to the vibronic model discussed in the previous subsection.
In these approaches one typically considers an infinite set of harmonic oscillators forming the bath.
It is easy to see that the Hamiltonian (\ref{HamTotExc}) can be written as
$H^e=H_{\rm sys}+H_{\rm sys-bath}+H_{\rm bath} $ with $H_{\rm sys}$ given by the purely electronic part (\ref{eq:H_elec}) with transition energies $\varepsilon_n$ that already include static overall shifts induced by the environment. The bath is taken as $H_{\rm bath}=\sum_n H^g_n$ and the system-bath coupling is described by a linear coupling of the electronic excitation of a monomer to the set of vibrational modes, i.e.\ $H_{\rm sys-bath}=\sum_n\ket{\pi_n}\bra{\pi_m}\otimes \sum_j \tilde\kappa_{nj} Q_{nj}$. The coupling constant $\tilde \kappa_{nj}$ describes the coupling of the excitation on monomer $n$ to the harmonic oscillator with frequency $\omega_{nj}$.
Typically, the frequency spectrum of bath oscillators is taken to be continuous so that the coupling constant becomes a continuous function of frequency.
The open system approaches have been used to describe optical and transfer properties of  light harvesting systems \cite{IsFl09_234111_,MoWuHu11_3045_} and can also be applied to explain properties of organic dye aggregates \cite{RoStEi10_5060_}.
We note that the HRS model can be considered as a special case of the open system models for which the bath is Markovian (i.e.\ memoryless).

\subsection{Mixed QM/MM}
With the increase of computer capabilities, in recent years it has become possible to simulate aggregates starting from a microscopic description \cite{ScTe91_421_,DaKoKl02_031919_,OlJaLi11_8609_,ShReVa12_649_}.
Since a full quantum mechanical treatment is still out of reach, one uses  mixed quantum classical approaches. In these approaches usually the nuclei are propagated classically in the electronic ground state using molecular dynamics. Along these nuclear trajectories one calculates the (time-dependent) transition energies using quantum chemistry methods. These time-dependent transition energies can then be used to obtain input parameters for open system models or to use them directly as the energy fluctuations in HSR model.
However, both approaches have their problems, since one is restricted to a classical propagation in the electronic ground state.

\subsection{Semi-empirical approach based on polarizabilities}
While the approaches discussed above start from a microscopic description of the exciton-phonon coupling and the environment, in the semi-empirical approaches one can take the standpoint that all the relevant information of the coupling of an electronic transition to other degrees of freedom is already contained in the shape of the absorption spectrum. The absorption spectrum in turn is linked to the polarizabilities of the monomers.

The induced electric moment of a monomer depends on the local field at that monomer, which is produced by the electric moments of all the other monomers and the external field. This leads to a set of coupled equations from which one can extract, for a given arrangement of the monomers, the polarizability of the aggregate and thus the optical properties. Such equations have been derived using classical treatments \cite{De64_393_}  or quantum approaches \cite{McBa64_581_,BrHe71_865_}. This method implicitly assumes that the frequency dependent polarizability does not change (beside an overall shift in energy) when going from the monomer to the aggregate.
Recently, it has been demonstrated that using measured monomer spectra as input  one can accurately describe the band-shape of J- and H-aggregates \cite{EiBr07_354_,EiBr06_376_}. This approach also gives an intuitive explanation of the vibrational structure and the broadenings of the absorption bands of the aggregate.

\subsection{Beyond the single-exciton approximation} Photonic
device applications may require structures operating in a
nonlinear response regime, where interactions between excitons
cannot be neglected. For example, thin J-aggregate films switches
based on optical bistability have been suggested in
\cite{MaGlFe00_1170_}. At higher intensities, however, an
additional loss-channel -- exciton-exciton annihilation --
appears. The underlying physical process is similar to the Auger
effect, followed by a fast non-radiative energy conversion. Within
an approximate picture this can be viewed as a process by which
two excitons localized on neighboring molecules are combined into
a higher-energy electronic excitation of a single molecule. Then,
the resulting excitation quickly decays to the lowest excited
state or ground state dissipating the exceeding energy through
vibrations. A detailed kinematic model for the exciton-exciton
annihilation in molecular crystals of different dimensions has
been introduced by Suna in \cite{Suna_PRB1970}, and a similar
master equation approach for exciton annihilation dynamics in
natural photosynthetic complexes has been suggested in
\cite{PaSw_BiophysJ79}. While the microscopic approaches discussed
above are still valid, the molecular Hamiltonian (\ref{eq:H_elec})
should be extended to account for at least two excited states per
each monomer \cite{KnSp_PRL95,Loch_Rev2011}.

\section{Hybrid excitonic structures}
\label{sec:Hybrid_structures}

In this section we specifically focus on J-aggregates. Their large absorption and fluorescence cross-sections combined with a narrow line width and a small Stokes shift allow for a coherent coupling between excitons and photons in optical cavities or excitons and plasmon modes in metal structures. This property can be utilized as an interface between photonic, plasmonic and excitonic circuits.

\subsection{Exciton-polariton structures}
\begin{figure}
  \begin{center}
   \includegraphics[width=1\linewidth, clip=true]{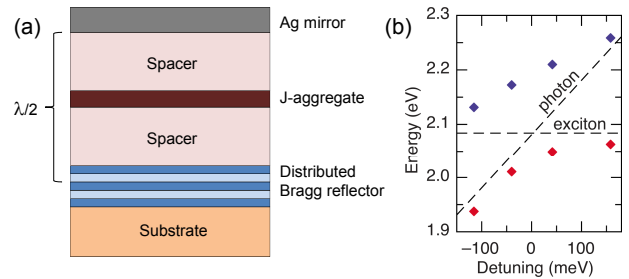}
  \caption{(a) Structure of a planar $\lambda/2$ organic microcavity with a thin layer of J-aggregates embedded. (b) An example of a cavity photon and a J-aggregate exciton energy level anticrossing. Reprinted with permission from Ref.~\cite{Akselrod2010}. Copyright 2010 by the American Physical Society.}
    \label{fig:cavity}
  \end{center}
\end{figure}

Most of the studies of strong coupling between excitons in J-aggreagtes and photons have been done using organic microcavities \cite{Skolnick_PRL99, Mahrt_ChemPhysLett01, Walker_APL02, Bulovic_PRL05}. A schematic illustration of an organic microcavity -- a planar $\lambda/2$ cavity with a J-aggregate layer embedded in it -- is shown in Fig.~\ref{fig:cavity}(a).
Strong coupling of cavity photons with excitons in a J-aggregate result in formation of exciton-polariton modes. This is associated with the splitting of a cavity mode -- vacuum Rabi splitting, see Fig.~\ref{fig:cavity}(b).

Within a simple model assuming only one exciton state and a single photon mode the value of the Rabi splitting, $\Omega_{\rm R}$, is just
\begin{equation}
\Omega_{\rm R} = \vec{d}_{\rm X} \cdot \vec{E}_{\rm vac},
\label{Rabi_splitting}
\end{equation}
 where $\vec{d}_{\rm X}$ is the transition dipole of the narrow exciton transition, $\vec{E}_{\rm vac}$ is the vacuum field in the position where the J-aggregate is located, and we assume that the vacuum field does not vary substantially on the scale of the aggregate. In J-aggregates the value of the exciton transition dipole scales with the number $N^*$ of coherently coupled molecules involved as $\vec{d}_{\rm X} \sim \sqrt{N^*}$. The value of $N^*$ is usually much smaller than $N$ - the total number of molecules in the aggregate, since the environment leads to localisation of the exciton states. The value of the Rabi splitting observed for J-aggregates strongly coupled with optical cavities ranges from several tens of meV \cite{Skolnick_PRL99, Mahrt_ChemPhysLett01} to several hundreds of meV \cite{Walker_APL02, Bulovic_PRL05, Akselrod2010} depending on the disorder present in the aggregate and the design of the cavity. Therefore, the splitting can be observed even at room temperatures. The photoluminescence in these structures is usually observed from the lower polariton branch only, due to fast relaxation of polaritons. By detunning the frequency of the cavity mode from the exciton resonance transition one could control the mixture of the exciton and photon and therefore modify its coherence properties making it more photon-like or exciton-like. Electroluminescence from a polariton mode has been demonstrated using a light-emitting-diode structure with a J-aggregate layer \cite{Bulovic_PRL05}. Moreover, lasing from exciton-polariton mode has been shown for cavities with single anthracene crystals \cite{Forrest2010} and some preliminary results were reported on lasing from cavities with cyanine dye J-aggregates \cite{Gleb_Telluride}. Large interest has been attracted to the idea of polariton Bose-Einstein condensation (BEC) in organic cavities. This may bring the rich and controversial quantum physics of non-equilibrium polariton condensates \cite{Snoke_BECBook, Deng_RMP2010,BuKa_NPhot12,Dev_NPhot12} observed at low temperatures in inorganic systems \cite{BEC_Nat06, BEC_NatPhys08, BEC_Nat09} up to room temperatures. However, to the best of our  knowledge no confirmed observations of BEC in organic cavities have been reported yet.

\subsection{Plexciton structures}
A strong coupling of excitons in molecular aggregates with plasmons in noble metals has been demonstrated for both propagating plasmon modes in films \cite{Bellessa2004} as well as localized modes in various types of nanostructures \cite{Sugawara_PRL06, Naomi_NanoLett08} and nanostructure arrays \cite{Bellessa_PRB09}.The hybrid plasmon-exciton modes were also named plexcitons \cite{Naomi_NanoLett08}. The observed values of the splitting between the plexciton modes are of the order of several hundreds of meV. The linewidth of plasmon modes is usually sufficiently larger than the linewidth of excitons in molecular aggregates (plasmon lifetime is about tens of femtoseconds as compared to picosecond exciton lifetime in J-aggregates). Therefore, the interaction between the exciton and the plasmon modes  can frequently be considered as the coupling of a single mode (exciton) to a broader continuum (plasmon). This results in the formation of a Fano resonance \cite{Fano, Govorov_PRL06}. It is interesting to notice that in the case of a strong optical pumping of plexcitonic structures non-linear Fano effects can be observed \cite{Govorov_NanoLett11}

\section{Conclusions}
\label{sec:Conclusions}

Aggregates of organic molecules -- supramolecular assemblies with strong resonant near-field interactions between electronic transitions -- could be exploited in the design of nanophotonic devices at the true nanometer scale.
The molecules, forming the aggregates, interact collectively with optical fields, and the absorbed energy is transferred in the form of excitons on a sub-micron scale. The exciton transport within the aggregates possesses coherent properties even at room temperatures, similar to the electromagnetic wave propagation through a medium. Moreover, molecular aggregates can be coupled coherently to photonic and plasmonic structures. While at the present stage there is a sufficient gap between the research communities studying excitonics and photonics, this review calls for merging the knowledge from the two fields.

\acknowledgements
The authors thank Gleb Akselrod and Brian Walker for providing experimental spectra. We also appreciate comments from Alex Govorov on hybrid structures. The Harvard contribution of this work was supported by the Defense Threat Reduction Agency under Contract No HDTRA1-10-1-0046. S. V. acknowledges support from the Center for Excitonics, an Energy Frontier Research Center funded by the U.S. Department of Energy, Office of Science and Office of Basic Energy Sciences under Award Number DE-SC0001088 as well as support from the Defense Advanced Research Projects Agency under award number N66001-10-1-4060.

\bibliographystyle{journal_v5_titles}
\bibliography{Refs_NP}

\end{document}